\documentclass[pre,twocolumn,superscriptaddress,showpacs,preprintnumbers,
eqsecnum,amsmath,amssymb]{revtex4}

\usepackage{graphicx}

\begin{document}

\title{Statistical wave scattering through classically chaotic cavities in the presence of surface absorption}

\author{M. Mart\'{\i}nez-Mares}
\affiliation{Departamento de F\'{\i}sica, Universidad Aut\'onoma
Metropolitana-Iztapalapa, 09340 M\'exico D. F., M\'exico}

\author{P. A. Mello}
\affiliation{Instituto de F\'{\i}sica, Universidad Nacional Aut\'{o}noma de
M\'{e}xico, 01000 M\'{e}xico D. F., M\'exico.}

\date{\today}

\begin{abstract}
We propose a model to describe the statistical properties of wave scattering through a classically chaotic cavity in the presence of surface absorption.
Experimentally, surface absorption could be realized by attaching an ``absorbing patch" to the inner wall of the cavity. In our model, the cavity is connected to the outside by a waveguide with $N$ open modes (or channels), while an experimental patch is simulated by an ``absorbing mirror" attached to the inside wall of the cavity; the mirror, consisting of a waveguide that supports $N_a$ channels, with absorption inside and a perfectly reflecting wall at its end, is described by a subunitary scattering matrix $S_a$. The number of channels $N_a$, as a measure of the geometric cross section of the mirror, and the lack of unitarity 
$P_a =\openone _{N_a} - S_a^{\dagger}S_a$, as a measure of absorption, are under our control: these parameters have an important physical significance for real experiments. The absorption strength in the cavity is quantified by $\gamma_a=\text{tr}P_a$. The statistical distribution of the resulting $S$ matrix
for $N=1$ open channel and only one absorbing channel, $N_a =1$, is solved
analytically for the orthogonal and unitary universality classes, $\beta =1$ and
$\beta =2$, respectively, and the results are compared with those arising from numerical simulations. The relation with other models existing in the literature, in some of which absorption has a volumetric character, is also studied.
\end{abstract}


\maketitle

%--------------------------------------------------------------------++

\section{Introduction}
\label{sec:intro}

Wave scattering experiments with microwave \cite{Richter,Stoekmann}
and acoustic resonators \cite{Schaadt} represent a fruitful field for the
verification of Random Matrix Theory (RMT) predictions \cite{Mello2004,LesHouches,Metha}
since, in that experimental domain, external parameters are particularly easy to control. However, the cost to be paid is the presence of absorption due to power
loss in the walls of the device used in the experiments.
Since the appearance of the paper by Doron {\it et al} \cite{Doron1990} which showed the drastic influence of absorption, many investigations, both experimental and theoretical, have been devoted to the study of absorption effects on transport properties \cite{Brouwer1997,Kogan2000,Lewenkopf1992,Beenakker2001,Schanze2001,
Schafer2003,Savin2003,Mendez-Sanchez2003,Fyodorov2003,Fyodorov2004,Savin2004,
Hemmady2004,Schanze2004,Kuhl2004,Savin2005} of classically chaotic cavities.

A simple model to describe the statistical properties of cavities including absorption was proposed by Kogan {\it et al.} \cite{Kogan2000}: it describes the system through a subunitary scattering matrix $S$, whose statistical distribution satisfies a maximum information-entropy criterion. The model turns out to be valid only in the strong-absorption limit. An alternative model proposed by Lewenkopf {\it et al.} \cite{Lewenkopf1992} simulates absorption by means of $N_p$ equivalent
``parasitic channels", not directly accessible to experiment (in addition to the $N$ physical channels), each one having an imperfect coupling to the cavity described by the transmission coefficient $T_p$. These parasitic channels can be interpreted in terms of the voltage-probe model originally proposed by B\"uttiker \cite{Buettiker1986,Buettiker1988}, where a fictitious lead with $N_p$ channels is attached to the cavity, each one with a coupling $T_p$ \cite{Brouwer1997}. The total scattering matrix $\hat{S}$ for this problem is unitary and has dimension $N+N_p$; the $N$-dimensional submatrix $\tilde{S}$ thereof is the physical scattering matrix which is of course subunitary. In the limit $N_p\rightarrow\infty$ and $T_p\rightarrow 0$, while the product $\gamma_p=N_pT_p$ is kept fixed and interpreted as the absorption strength, this model was shown by Brouwer and Beenakker \cite{Brouwer1997} to describe {\em volume absorption},
in the sense that the problem described by $\tilde{S}$ is equivalent to one in which
all the energy levels of the closed cavity acquire a fixed imaginary part, associated precisely with the absorption strength. Although the above quantity $\gamma$ is not directly under experimental control, it can be chosen so as to fit the experimental data \cite{Schanze2004,Kuhl2004}.

Alternatively, instead of considering a cavity that exhibits volume absorption,
we may think of an experimental situation in which ``absorbing patches"
are literally attached to the inner wall of the cavity in a controllable fashion, 
giving rise to what we shall call {\em surface absorption}. The physical properties of these patches could be determined and controlled independently of the cavity they are used with.

\begin{figure}
\includegraphics[width=5.0cm]{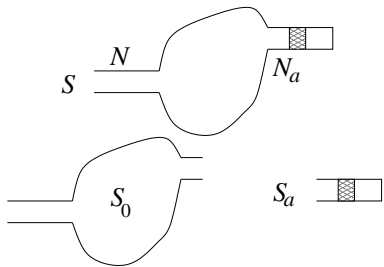}
\caption{A ballistic chaotic cavity described by the scattering matrix $S_0$,
connected to a waveguide with $N$ propagating
modes and to an ``absorbing mirror". The latter is formed by
a ``frustrated waveguide" that supports $N_a$ channels and contains
an absorbing barrier inside and a perfectly reflecting mirror at its end;
the absorbing mirror is described by the subunitary scattering matrix $S_a$.}
\label{fig:model}
\end{figure}

In the present paper we propose a simple model to describe the statistical properties of wave scattering through a ballistic chaotic cavity in the presence of surface absorption as defined above.  In the model [see Fig. \ref{fig:model}], the cavity is connected to the outside by a waveguide with $N$ open modes (or channels),
while an experimental patch is simulated by attaching to the inside wall of the cavity an ``absorbing mirror", consisting of a ``frustrated waveguide" that supports $N_a$ channels, with absorption inside and a perfectly reflecting wall at its end: the mirror is described by a subunitary scattering matrix $S_a$.
The number of channels $N_a$, as a measure of the geometric cross section of the mirror, and the lack of unitarity $P_a =\openone _{N_a} - S_a^{\dagger}S_a$, as a measure of absorption, are under our control: these parameters have an important physical significance for real experiments. The absorption strength in the cavity is quantified by $\gamma_a=\text{tr}P_a$. Alternatively, we may think of attaching to the inner wall of the cavity $N_a$ absorbing patches, each one with a geometric cross section that supports only one channel with a known absorption.

The paper is organized as follows. In the next section we study the model for surface absorption that was described in the previous paragraph.
In Sect. \ref{sec:voltage-probe} we find the relation between the present model and the parasitic-channel model also described above: we show the equivalence ot the two models when, in the latter, the limit mentioned earlier leading to volume absorption is not taken and the parameters are chosen properly.
In Sect. \ref{sec:one-channel} we show that the model can be solved analytically in the one-channel case and for one absorbing channel.
Finally, we present our conclusions in Sec. \ref{sec:conclusions}. 

\section{The model for surface absorption}
\label{sec:surface-model}

The system that we are interested in and that we shall study in the present section was described in the Introduction and is shown in Fig. \ref{fig:model}. 

The scattering problem can be studied in terms of an $N\times N$ scattering matrix $S$, which in the stationary case relates the outgoing-wave to the incoming-wave
amplitudes \cite{Mello2004,Newton1982}. As shown schematically in Fig. \ref{fig:model}, $S$ can be seen as the combination of a scattering matrix $S_0$ which describes a ballistic cavity connected to two non-absorbing waveguides
(one with $N$ channels on the left and the other with $N_a$ channels on the right)
and the scattering matrix $S_a$ that describes the absorbing mirror; $S_a$ is a fixed $N_a\times N_a$ subunitary matrix which can be considered as an input to the problem. The lack of unitarity of $S_a$ is defined by $P_a=\openone_{N_a}-S_a^{\dagger}S_a$: in the most general situation, $S_a$, and hence $P_a$, is not a diagonal matrix. The overall absorption strength of the absorbing mirror can be quantified by $\gamma_a=\text{tr} P_a$.

The scattering matrix $S_0$ is of dimension $N+N_a$ and has the structure
\begin{equation}
S_0 = \left(
\begin{array}{cc}
r_0 & t_0' \\ t_0 & r_0'
\end{array}
\right) ,
\label{S0}
\end{equation}
where $r_0$ is the $N\times N$ reflection matrix for incidence on the left, $r_0'$ is the $N_a\times N_a$ reflection matrix for incidence on the right and $t_0$ and $t_0'$ are the corresponding transmission matrices of dimensions $N_a\times N$ and
$N\times N_a$, respectively.

The matrix $S_0$ will be considered to belong to one of the basic symmetry classes introduced by Dyson in Quantum Mechanics \cite{Dyson1962}, which we briefly recall in order to make a more precise statement. In the ``unitary'' case, also denoted by $\beta=2$, the only restriction on $S_0$ is unitarity, due to the physical requirement of flux conservation. In the ``orthogonal'' case ($\beta=1$) $S_0$ is also symmetric because of either time-reversal invariance (TRI) and integral spin, or TRI, half-integral spin and rotational symmetry. In the ``symplectic case ($\beta =4$), $S_0$ is self-dual because of TRI with half-integral spin and no rotational symmetry. In the scattering problem of scalar classical waves, the orthogonal case is the physically relevant one. However, we shall consider below both $\beta =1$ and $\beta =2$, in the understanding that the unitary case is to be considered as a reference problem, as it is often simpler to treat mathematically than the orthogonal one.

We assume that the barrier inside the absorbing mirror is sufficiently far apart from both the entrance to the cavity at one end and the perfect reflector at the other end that the evanescent modes do not play a role \cite{Mello2004}.
Then, following the combination rule of scattering matrices, the total $S$ matrix which describes the cavity with absorption is a subunitary matrix given by
\begin{equation}
S = r_0 + t_0' \frac{1}{\openone_{N_a} - S_a r_0'} S_a t_0 \; ,
\label{Smmm}
\end{equation}
where $\openone_n$ stands for the unit matrix of dimension $n$.

The physics of Eq. (\ref{Smmm}) is clear. The first term on the right hand side is the reflected part of the wave that enters the cavity; the transmitted part reaches the absorbing mirror with the (matrix) amplitude $t_0$ and the fraction $S_a$ (again a matrix) thereof returns to the cavity; the wave then suffers a multiple scattering process between the cavity and the absorbing mirror, until finally the fraction $t_0'$ leaves the cavity through the lead on the left.

The statistical features of the quantum mechanical scattering produced by the cavity
shown in the lower portion of Fig. \ref{fig:model} --assumed to have a chaotic classical dynamics-- are described by a measure in the $S_0$-matrix space which, through the assumption of ergodicity, gives the probability of finding $S_0$ in a given volume element as the energy $E$ changes and $S_0$ wanders through that space
\cite{Mello2004}. In the absence of direct processes, or short paths, $S_0$ is taken to have a uniform distribution, given by the invariant measure for the symmetry class in question. The invariant measure, denoted by $d\mu_{\beta}(S_0)$, is defined by its invariance under the symmetry operation relevant to that universality class \cite{Dyson1962,Hua}; i.e.,
\begin{equation}
d\mu_{\beta}(S_0) = d\mu_{\beta}(U_0S_0V_0),
\label{invmeasure}
\end{equation}
where $U_0$ and $V_0$ are arbitrary but fixed unitary matrices in the unitary case,
while $V_0=U_0^T$ in the orthogonal case. This defines the circular unitary
(orthogonal) ensemble, CUE (COE) for $\beta=2$ ($\beta=1$).

The distribution
\begin{equation}
dP_{\beta }(S)=p_{\beta }(S)d\mu_{\beta }(S)
\label{p(S)}
\end{equation}
of the resulting scattering matrix $S$ of Eq. (\ref{Smmm}) for the full system consisting of the cavity plus the absorbing mirror can in principle be calculated for a given $S_a$, assuming that $S_0$ is distributed according to the invariant measure, Eq. (\ref{invmeasure}). This procedure is particularly well suited for numerical simulations. Indeed, it is used in Sec. \ref{sec:one-channel} to study numerically the case of one open channel ($N=1$) and one absorbing channel ($N_a=1$).

One property of $p_{\beta }(S)$ is immediately obvious: the average of $S$ vanishes, i.e., $\langle S\rangle=0$, as we can see by expanding the right-hand side
of Eq. (\ref{Smmm}) and averaging term by term.

Although the explicit expression for $p_{\beta }(S)$ is not known for an arbitrary number of channels, the case of one open channel ($N=1$) and one absorbing channel ($N_a=1$) can be solved analytically: this is done in Sec. \ref{sec:one-channel}, exploiting the relation with the parasitic-channel model that is explained in the next section.

\section{Relation between the surface-absorption model and the parasitic-channel model}
\label{sec:voltage-probe}

\begin{figure}
\includegraphics[width=4.0cm]{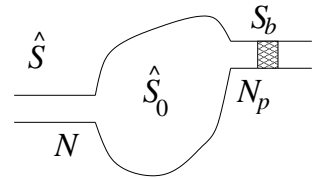}
\caption{In the parasitic-channel model, a ballistic chaotic cavity connected to a waveguide with $N$ propagating modes is also connected to an additional, fictitious, waveguide supporting $N_p$ ``parasitic" channels. Inside the latter waveguide there is a (non-absorbing) barrier described by the scattering matrix $S_b$. Only the $N\times N$ submatrix $\widetilde{S}$ of the total scattering matrix of this system is accessible to experiment; it is subunitary and is taken as a model for a real cavity in the presence of absorption.}
\label{fig:parasitic-channel}
\end{figure}

In the parasitic-channel model described in the Introduction and sketched in Fig. \ref{fig:parasitic-channel} the cavity is attached to a waveguide supporting $N$ physical channels, while absorption is simulated by attaching an additional, fictitious, waveguide supporting $N_p$ equivalent ``parasitic'' channels, each one with a coupling given by a transmission coefficient $T_p$ \cite{Lewenkopf1992,Brouwer1997}. 

We denote by $\widehat{S}$ the scattering matrix that describes the whole system
shown in Fig. \ref{fig:parasitic-channel}, including the fictitious waveguide and
the (non-absorbing) barrier inside it; $\widehat{S}$ is of dimension $N+N_p$ and has the structure
\begin{equation}
\widehat{S}=\left(
\begin{array}{cc}
\widetilde{S} & \widehat{S}_{1p} \\ \widehat{S}_{p1} & \widehat{S}_{pp}
\end{array}
\right) ,
\label{Shat}
\end{equation}
where the indices $1$ and $p$ denote the set of $N$ physical and $N_p$ fictitious channels, respectively; it is unitary for $\beta=2$ and unitary-symmetric for $\beta=1$. The $N\times N$ submatrix $\widetilde{S}$ describes the cavity in the presence of absorption, as one would observe it in an experiment in which only the physical waveguide were present; $\widetilde{S}$ is subunitary and will be referred to as the ``physical" $S$ matrix.

The tunnel barrier in the fictitious waveguide in Fig. \ref{fig:parasitic-channel},
described by the  $2N_p \times 2N_p$ unitary scattering matrix $S_b$
\begin{equation}
S_b 
= \left(
\begin{array}{cc}
\sqrt{1-T_p}\, \openone_{N_p}  &  i\sqrt{T_p}\, \openone_{N_p}  \\
i\sqrt{T_p}\, \openone_{N_p}  & \sqrt{1-T_p}\, \openone_{N_p}
\end{array}
\right) \; ,
\label{Sb EC}
\end{equation}
implies an imperfect coupling of the parasitic channels to the cavity; it can also be interpreted as giving rise to direct processes in a system described by $\widehat{S}$ \cite{Brouwer1995-7}. Then, for a cavity whose classical dynamics is chaotic, $\widehat{S}$ is distributed according to the Poisson kernel
\cite{Mello1985,Friedman1985,Mello1999}
\begin{equation} \label{Pkernel}
dP_K^{(\beta)}(\widehat{S}) =
\frac{ \left[ \det \left( \openone_{N_T} - \langle \widehat{S} \rangle
\langle \widehat{S} \rangle^{\dagger} \right) \right]^{(\beta N_T+2-\beta)/2} }
{\left| \det \left( \openone_{N_T} - \widehat{S}
\langle \widehat{S} \rangle^{\dagger} \right)
\right|^{\beta N_T+2-\beta} } d\mu_{\beta}(\widehat{S}) ,
\end{equation}
where $N_T=N+N_p$, $d\mu_{\beta}(\widehat{S})$ is the invariant measure for
$\widehat{S}$ defined as in Eq. (\ref{invmeasure}), assumed to be normalized
to unity, and
\begin{equation}
\langle \widehat{S} \rangle = \left(
\begin{array}{cc}
0_N & 0 \\ 0 & \sqrt{1-T_p} \openone_{N_p}
\end{array}
\right) \, .
\label{<S> EC}
\end{equation}
The probability distribution of the physical scattering matrix $\widetilde{S}$ that we are interested in has to be calculated from Eq. (\ref{Pkernel}). This is how Brouwer and Beenakker showed that, in the limit $N_p\rightarrow\infty$ and $T_p\rightarrow 0$, with the product $\gamma_p=N_pT_p$ being kept fixed and interpreted as the absorption strength, this model describes volume absorption, in the sense that the problem represented by $\tilde{S}$ is equivalent to one in which
all the energy levels of the closed cavity acquire a fixed imaginary part, 
associated precisely with that absorption strength 
\cite{Brouwer1997,Beenakker2001}.

Let us now think of the full system shown in Fig. \ref{fig:parasitic-channel}
as the combination of the cavity connected to the two waveguides (in the absence of the barrier) and described by the $(N+N_p)\times(N+N_p)$ scattering matrix
\begin{equation}
\widehat{S}_0 = \left(
\begin{array}{cc}
\hat{r}_0 & \hat{t}_0' \\ \hat{t}_0 & \hat{r}_0'
\end{array}
\right) \; ,
\label{Shat0}
\end{equation}
plus the barrier described by $S_b$. We shall {\em not} take, in what follows, the limit mentioned in the previous paragraph, i.e., $N_p$ will be kept finite.
As we shall see, it will also be convenient, for the description of the barrier,
to drop the ``equivalent-channel" assumption normally made in this context [Eq. (\ref{Sb EC})] and consider a barrier matrix $S_b$ with the general structure
\begin{equation}
S_b =
\left(
\begin{array}{cc}
r_b & t'_b \\
t_b & r'_b
\end{array}
\right) \; .
\label{Sb}
\end{equation}
The total scattering matrix $\widehat{S}$ is then given by
\begin{equation}
\widehat{S}
= r_B + t'_B \frac{1}{\openone_{N+N_p} - \widehat{S}_0 r'_B}\widehat{S}_0 t_B \; .
\label{Shat 1}
\end{equation}
Here we have introduced the $(N+N_p)\times(N+N_p)$ matrices
\begin{subequations}
\begin{eqnarray}
r_B &=&\left(
\begin{array}{cc}
0_{N} &   0 \\
0     &   r_b
\end{array}
\right)  ,
\hspace{.5cm}
t'_B =\left(
\begin{array}{cc}
\openone_{N} &   0 \\
0     &   t'_b
\end{array}
\right) ,
\label{SB 1}
\\
t_B &=&\left(
\begin{array}{cc}
\openone_{N} &   0 \\
0     &   t_b
\end{array}
\right)  ,
\hspace{.5cm}
r'_B =\left(
\begin{array}{cc}
0_{N} &   0 \\
0     &   r'_b
\end{array}
\right) ,
\label{SB 2}
\end{eqnarray}
\label{SB}
\end{subequations}
which denote the barrier reflection and transmission matrices associated with
the {\em two} waveguides in Fig. \ref{fig:parasitic-channel} [notice that in the present model the left waveguide has no barrier; hence the matrices $0_{N}$ and $\openone_{N}$ in Eq. (\ref{SB}); in a more complete model there might be a barrier in the left waveguide too (see comment at the end of Sec. \ref{sec:one-channel})].
The physical $S$ matrix $\tilde{S}$ of Eq. (\ref{Shat}) can then be found from Eq. (\ref{Shat 1}) to be
\begin{equation}
\widetilde{S}
= \hat{r}_0 + \hat{t}_0'
\frac{1}{\openone_{N_p} - r'_b \, \hat{r}_0'}\; r'_b \; \hat{t}_0 \; .
\label{Svp}
\end{equation}

Comparing Eqs. (\ref{Smmm}) and (\ref{Svp}) we see that the model for surface absorption of Sec. \ref{sec:surface-model} and the parasitic-channel model of the present section can be made equivalent, in the sense that $\widetilde{S}=S$,
choosing the number of parasitic channels in the latter to coincide with the number of absorbing channels in the former, i.e., $N_p=N_a$, choosing the two cavities to be identical, i.e., $\widehat{S}_0=S_0$, and the barrier reflection matrix $r'_b$ of the present section to coincide with the absorption matrix of Sec. \ref{sec:surface-model}, i.e., $r'_b =S_a$. 

Within the equivalent-channel assumption, equivalence of the two models would require $S_a = \sqrt{1-T_p}\, \openone_{N_p}$. If in our surface-absorption model we were to take the limit that was described right after Eq. (\ref{<S> EC}) in connection with the parasitic-channel model, we would obtain the same answer as for volume absorption by identifying $\gamma _p$ with $\gamma_a=\text{tr} P_a$ where $P_a=\openone_{N_a}-S_a^{\dagger}S_a$.

The statistical distribution of the scattering matrix $\widehat{S}$ of Eq. (\ref{Shat 1}) is of course given by Poisson's kernel of  Eq. (\ref{Pkernel}), 
where the average $S$ matrix $\langle \widehat{S} \rangle$ is now given by
$\langle \widehat{S} \rangle = r_B$ [see Eq. (\ref{SB 1})], instead of Eq. (\ref{<S> EC}). Using the identification described right after Eq. (\ref{Svp}),
the calculation of the resulting distribution of $\tilde{S}$ also gives the distribution of $S$, Eq. (\ref{Smmm}), for the model of the previous section.
This procedure will be taken advantage of in the next section, in order to analyze the case $N=1$, $N_a =1$.

\section{Analytical solution of the surface-absorption model for the case $N=1$, $N_a=1$}
\label{sec:one-channel}

In this section we calculate, within the surface-absorption model of Sec. \ref{sec:surface-model}, the distribution of the scattering matrix $S$ of Eq. (\ref{Smmm}) which describes a chaotic cavity perfectly connected to a waveguide supporting one channel ($N=1$), and in the presence of one absorption channel ($N_a=1$); $S$  is thus a complex number that we parametrize as
\begin{equation}
S = \sqrt{R} \, e^{i\theta} .
\label{S-1}
\end{equation}
The absorbing mirror is also described by a subunitary matrix $S_a = \sqrt{R_a} \, e^{i\theta_a}$.

We can take advantage of the equivalence with the parasitic-channel model discussed in the previous section, with the identification: $N_p=N_a=1$ and $T_p=P_a=1-R_a$, $\theta _a =0$. For the present case, $\widehat{S}$ of Eq. (\ref{Shat}) is a $2\times 2$ unitary matrix for $\beta=2$, with the additional condition of symmetry for $\beta=1$. It can be parametrized in a polar representation as \cite{Mello1988,Mello1991}
\begin{equation}
\widehat{S} =
\left[
\begin{array}{cc}
-\sqrt{1-\tau} e^{i(\phi+\phi')} &
\sqrt{\tau} e^{i(\phi+\psi')} \\
\sqrt{\tau} e^{i(\phi'+\psi)} &
\sqrt{1-\tau} e^{i(\psi+\psi')}
\end{array}
\right],
\label{S par}
\end{equation}
where $0\le\phi,\psi,\phi',\psi'<2\pi$ and $0\le\tau\le 1$; for $\beta=1$ we have the restrictions $\phi'=\phi$ and $\psi'=\psi$. The 11 matrix element of $\widehat{S}$ is the physical 
$\widetilde{S}\equiv S=-\sqrt{1-\tau}e^{i(\phi+\phi')}$, so that the parameters of $S$ appearing in (\ref{S-1}) are given by
\begin{subequations}
\begin{eqnarray}
R &=& 1-\tau \\
\theta &=& \phi+\phi'+\pi.
\end{eqnarray}
\end{subequations}
The matrix $\widehat{S}$ is distributed according to Eq. (\ref{Pkernel}), with 
$N_T=N + N_p=2$ and
\begin{equation}
\langle \widehat{S} \rangle = \left(
\begin{array}{cc}
0 & 0 \\ 0 & \sqrt{R_a}
\end{array}
\right).
\label{Savehat-1}
\end{equation}
In the polar representation of Eq. (\ref{S par}), the invariant measure $d\mu_{\beta}(\widehat{S})$ needed in Eq. (\ref{Pkernel})
is given by \cite{Baranger1994,Jalabert1994}
\begin{equation}\label{invmeasS}
d\mu_{\beta}(\widehat{S})= \frac{d\tau}{(2\sqrt{\tau})^{2-\beta}}
\frac{d\phi}{2\pi} \frac{d\psi}{2\pi}
\left( \frac{d\phi'}{2\pi} \frac{d\psi'}{2\pi} \right)^{\beta-1}.
\end{equation}
The quantity
$\left| \det \left( \openone_{N + N_p} - \widehat{S}
\langle \widehat{S} \rangle^{\dagger} \right)
\right|$
appearing in Eq. (\ref{Pkernel}), i.e.,
\begin{eqnarray}
&&\left|
\det \left(
\openone_{2} - \widehat{S}\langle \widehat{S} \rangle^{\dagger}
\right)
\right|
\label{det} \nonumber \\
&& \;\;\;\;\; = 1+(1-\tau )R_a - 2\sqrt{1-\tau }\sqrt{R_a}\cos (\psi + \psi') ,
\nonumber \\
\end{eqnarray}
is independent of the angles $\phi, \phi '$ and hence of the phase $\theta $
of $S$, which is thus uniformly distributed. The normalized probability distribution of $S$ can thus be written as
\begin{equation}\label{ProbS}
dP_{\beta}(S) = P_{\beta}(R) \, dR \, \frac{d\theta}{2\pi}.
\end{equation}
What remains to be calculated is thus the probability distribution of the reflection coefficient $R$, $P_{\beta}(R)$. Refs. \cite{Mello1999} and \cite{Mello2004} give the distribution of $\tau $ for certain particular cases of the average $S$ matrix, of which Eq. (\ref{Savehat-1}) is a particular case.
In Eqs. (5.8) and (5.11) of Ref. \cite{Mello1999} (or Eqs. (6.81) and (6.85) of Ref. \cite{Mello2004}), we make the substitution $1-T=R$, together with $X=0$, $Y=\sqrt{R_a}$ in (5.8) and $X=\sqrt{R_a}$, $Y=0$ in (5.11) of Ref. \cite{Mello1999}, (or Eqs. (6.81) and (6.85) of Ref. \cite{Mello2004}, respectively), with the result
\begin{equation}\label{P(R)-2}
P_2(R) = \frac{(1-R_a)^2 (1+R_a R)}{(1-R_a R)^3}
\end{equation}
for $\beta=2$, and
\begin{equation}\label{P(R)-1}
P_1(R) = \frac{ (1-R_a)^{3/2} }{ 2\sqrt{1-R} } \,
{}_2 F_1 (3/2,3/2;R_a R)
\end{equation}
for $\beta=1$, ${}_2 F_1$ being a hypergeometric function \cite{Abramowitz}.

In principle, the distribution of $S$ could also be obtained from its definition,
Eq. (\ref{Smmm}), and the fact that $S_0$ is distributed according to the invariant
measure. This procedure is especially suited for numerical simulations.
The absorbing mirror is described by the $1\times 1$ subunitary scattering matrix
$S_a=\sqrt{R_a}e^{i\theta_a}$, where the phase $\theta_a$ and the reflection coefficient $R_a$ are fixed; $0\le R_a \le 1$ and for simplicity we may take, as above, $\theta_a=0$. The lack of unitarity of $S_a$ is given by $P_a=1-R_a$.
The matrix $S_0$ of Eq. (\ref{S0}) is a $2\times 2$ unitary matrix that can be parametrized in a polar representation as in Eq. (\ref{S par}), with the various parameters now having an index 0 and the range of variation: $0\le\phi'_0,\psi'_0,\phi''_0,\psi''_0<2\pi$ and $0\le\tau_0\le1$;
for $\beta=1$ we have the restrictions $\phi'_0=\phi_0$ and $\psi'_0=\psi_0$.
For a chaotic cavity, $S_0$ is distributed according to the invariant measure which
in terms of the polar parameters is given by
\begin{equation}\label{invmeasS0}
d\mu_{\beta}(S_0)= \frac{d\tau_0}{(2\sqrt{\tau_0})^{2-\beta}}
\frac{d\phi_0}{2\pi} \frac{d\psi_0}{2\pi}
\left( \frac{d\phi'_0}{2\pi} \frac{d\psi'_0}{2\pi} \right)^{\beta-1}.
\end{equation}
The distribution of $S$ can then be obtained from Eq. (\ref{Smmm}).

\begin{figure}
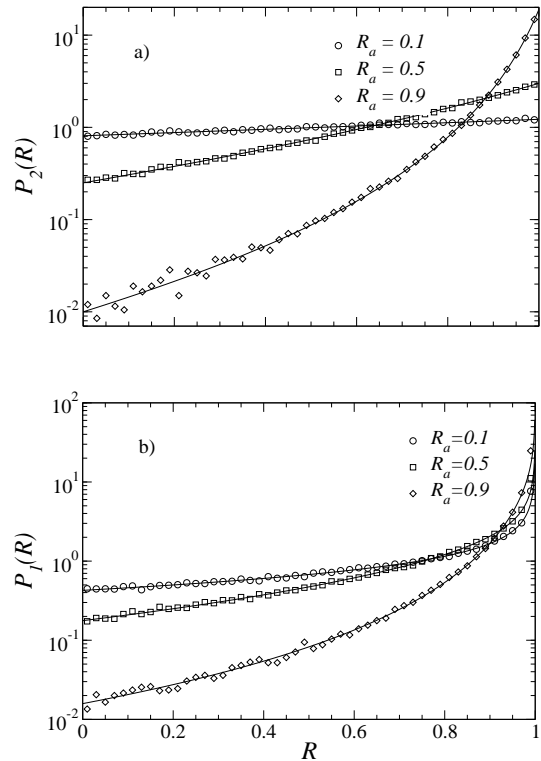

\includegraphics[width=7.0cm]{fig3.eps}
\vskip 0.8cm
\includegraphics[width=7.0cm]{fig4.eps}
\caption{Distribution of $R$ of Eq. (\ref{S-1}) for a) $\beta=2$ and b) $\beta=1$
symmetries, for one open channel, $N=1$, one absorbing channel, $N_a=1$, and $R_a=0.1$, 0.5, and 0.9. The continuous lines are the analytical results given by Eqs. (\ref{P(R)-2}) for $\beta=2$ and (\ref{P(R)-1}) for $\beta=1$. The numerical simulation was performed as explained in the text using $10^4$ samples; the different values of $R_a$ are distinguished by the symbols also shown in the figure. The agreement between theoretical and numerical results is excellent.}
\label{fig:Rdist}
\end{figure}

In Fig. \ref{fig:Rdist} we compare the analytical results of Eqs. (\ref{P(R)-2})
and (\ref{P(R)-1}) with the results of a numerical simulation obtained from the ensemble of Eq. (\ref{invmeasS0}) and the relation (\ref{Smmm}) between $S$ and $S_0$: the agreement is excellent.

We should point out that in real experiments the waveguide is not perfectly coupled
to the cavity, as has been shown in Ref. \cite{Mendez-Sanchez2003}, thus giving rise to direct processes. In such cases, the distribution of $S$ must be modified by the procedure of Ref. \cite{Kuhl2004}.

\section{Summary and Conclusions}
\label{sec:conclusions}

We have presented a model to describe the statistical properties of wave scattering
in a ballistic chaotic cavity in the presence of surface absorption: it is suggested that this type of absorption be realized in the laboratory by attaching 
one or several ``absorbing patches" to the inner wall of the cavity. The model simulates one such absorbing patch by means of an absorbing mirror, whose physical characteristics, i.e., its cross section and its absorption properties, are, in principle, experimentally measurable.

We compare our model with the parasitic-channel model introduced by other authors
and find the conditions under which the two are equivalent. It is found in the literature that in the limit of a large number of parasitic channels weakly coupled to the cavity, the parasitic-channel model describes absorption uniformly distributed over the volume of the cavity; in this limit, volume and surface absorption give the same results.

Finally, we show that our model is analytically solvable for the case of one open channel, $N=1$, and one absorbing channel, $N_a=1$, for arbitrary absorption
strength. The results of numerical simulations for this case are in excellent agreement with theory. In principle, this problem is amenable to experimental observation.

\acknowledgments

One of the authors (PAM) wishes to express his gratitude to the Max Plank Institut f\"ur Physik Complexer Systeme, Dresden, Germany, for its hospitality during the final stages of the writing of the manuscript, as well as Conacyt, for financial support through Contract No. ... MMM thanks R. A. M\'endez-S\'anchez for useful discussions.

%%%%%%%%%%%%%%%%%%%%%%%%%%%%%%%%%%%%%%%%%%%%%%%%%%%%%%%%%%%%%%%%%%%%%%%

%%%%%%%%%%%%%%%%%%%%%%%%%%%%%%%%%%%%%%%%%%%%%%%%%%%%%%%%%%%%%%%%%%%%%++

\end{document}